\documentclass[10pt,preprint]{sigplanconf}

\usepackage{xcolor}
\usepackage{listings}
\usepackage{xspace}
\usepackage{booktabs}
\usepackage{color}
\usepackage{colortbl}
\usepackage[font=small,format=plain,labelfont=bf,up,textfont=it,aboveskip=3.5pt]{caption}
\usepackage{float}  
\usepackage{subcaption}
\usepackage{balance}
\usepackage{url}
\usepackage{courier}
\usepackage{mathptmx}
\usepackage{hyperref}
\newcommand{\system}{Sonata\xspace}
\definecolor{steel_blue}{RGB}{70, 130, 180}
\usepackage{amsmath}
\usepackage{pifont}
\usepackage{algorithm}
\usepackage{algpseudocode}
\usepackage{placeins}

\newcommand{\ie}{{\em i.e.}}
\newcommand{\eg}{{\em e.g.}}

\newcommand{\smartparagraph}[1]{\noindent{\bf #1}\ }

\definecolor{darkgreen}{rgb}{0.0, 0.65, 0.0}
\newcommand{\COMMENTS}{yes}
\ifthenelse{\equal{\COMMENTS}{yes}}{%
\newcommand{\ag}[1]{\textit{\textcolor{blue}{[arpit]: #1}}} % Arpit's notes
\newcommand{\nf}[1]{\textcolor{red}{(\textit{\textcolor{red}{NF: #1}} )}}
\newcommand{\mcnote}[1]{\textit{\textcolor{blue}{[marco]: #1}}} % Marco's comments
\newcommand{\rjh}[1]{\textit{\textcolor{darkgreen}{[rob]: #1}}} % Rob's notes
}
{
\newcommand{\ag}[1]{}
\newcommand{\lv}[1]{}
\newcommand{\rmnote}[1]{}
\newcommand{\rbnote}[1]{}
\newcommand{\nf}[1]{}
\newcommand{\mcnote}[1]{}
\newcommand{\rjh}[1]{}
}

\lstdefinelanguage{Python}
{keywords={MapBolt, ReduceBolt, >>, >+, +, &, push, if_, elif_, else, pop, mapInit, 
match, fwd, modify, set, mod, sample, sampleD, sampleS, $\triangleleft$, announce, 
withdraw, mapD, mapS, map, reduce, filter, filterD, filterS, runningReduceD, runningReduceS, 
distinct, toList, mapValues, countS, countByWindow, countByValueAndWindow, reduceByKey, window,
transform, map-init, update-metadata, update-headers, emit}, 
keywordstyle=\bfseries\ttfamily,
keywordstyle=[2]\ttfamily\bfseries,% for example
commentstyle=\ttfamily, % white comments
stringstyle=\ttfamily, % typewriter type for strings
identifierstyle=\ttfamily,
emph={}, 
emphstyle=\ttfamily\bfseries\color{red},
sensitive=true, alsoletter={0,1,2,3,4,5,6,7,8,9,-,>>,+,&,|,_},comment=[l][\footnotesize\sffamily\textbf]{\!}
}

\lstdefinelanguage{query2}
{keywords={MapBolt, ReduceBolt, >>, >+, +, &, push, if_, elif_, else, pop, mapInit, 
match, fwd, modify, set, mod, sample, sampleD, sampleS, $\triangleleft$, announce, 
withdraw, mapD, mapS, map, reduce, filter, filterD, filterS, runningReduceD, runningReduceS, 
distinct, toList, mapValues, countS, countByWindow, countByValueAndWindow, reduceByKey, window,
transform, map-init, update-metadata, update-headers, emit}, 
keywordstyle=\bfseries\ttfamily,
keywordstyle=[2]\ttfamily\bfseries,% for example
commentstyle=\ttfamily, % white comments
stringstyle=\ttfamily, % typewriter type for strings
identifierstyle=\ttfamily,
emph={join, trafficAnomalyIPs}, 
emphstyle=\ttfamily\bfseries\color{red},
sensitive=true, alsoletter={0,1,2,3,4,5,6,7,8,9,-,>>,+,&,|,_},comment=[l][\footnotesize\sffamily\textbf]{\!}
}

\lstdefinelanguage{Scala}%
  {morekeywords={abstract,case,catch,class,def,%
    do,else,extends,false,final,finally,%
    for,if,implicit,import,lazy,match,mixin,%
    new,null,object,override,package,%
    private,protected,requires,return,sealed,%
    super,this,trait,true,try,%
    type,val,var,while,with,yield},%+
otherkeywords={=,=>,<-,<\%,<:,>:,\#,@},%
   sensitive,%
   morecomment=[l]//,%
   morecomment=[n]{/*}{*/},%
   morestring=[b]",%
   morestring=[b]',%
   morestring=[b]""",%
  }[keywords,comments,strings]%

\lstdefinelanguage{p4}
{keywords=[2]{control, action, else, if, table, register, field_list, field_List, field_list_calculation},
 otherkeywords={register_read, modify_field, apply, bit_or, register_write, input, algorithm, width, instance_count, exact,lpm,reads,actions  ,bit_and, clone_ingress_pkt_to_egress, add_header}, 
sensitive=true, alsoletter={-,>>,+,&,|,_},
basicstyle=\color{red}\ttfamily,
keywordstyle=\color{darkgreen}\ttfamily,
keywordstyle=[2]\ttfamily\bfseries\color{steel_blue},% for example
commentstyle=\ttfamily, % white comments
stringstyle=\ttfamily, % typewriter type for strings
identifierstyle=\ttfamily,
alsoletter={0,1,2,3,4,5,6,7,8,9}
comment=[l][\footnotesize\ttfamily\color{red}]{\!}
}

\usepackage{lastpage}
\usepackage{adjustbox}
\usepackage{lipsum}% example text

\newcommand{\supsym}[1]{\raisebox{4pt}{{\footnotesize #1}}}
\newcommand{\et}{\supsym{$\dag$}}
\newcommand{\kst}{\supsym{$\diamond$}}
\newcommand{\ptn}{\supsym{$\star$}}
\newcommand{\ua}{\supsym{$\ddag$}}
\newcommand{\nks}{\supsym{$\triangle$}}

\date{}
\title{\LARGE{\system: Query-Driven Network Telemetry}}
\authorinfo{
{Arpit Gupta\ptn, Rob Harrison\ptn, Ankita Pawar\ua, R\"udiger Birkner\et,} \\ 
{Marco Canini\kst, Nick Feamster\ptn, Jennifer Rexford\ptn, Walter Willinger\nks}\\
\ptn\normalsize{Princeton University}~~~\ua\normalsize{Unaffiliated}~~~\et\normalsize{ETH Z\"{u}rich}~~~\kst\normalsize{KAUST}~~~\nks\normalsize{NIKSUN Inc.}\\
}
\begin{document}
\thispagestyle{empty}

\maketitle

%\AcmCopyright
%\ToAppear

\begin{sloppypar}

\begin{abstract}  
Operating networks depends on collecting and analyzing measurement data.
Current technologies do not make it easy to do so, typically because they
separate data collection (\eg, packet capture or flow monitoring) from
analysis, producing either too much data to answer a general question, or too
little data to answer a detailed question. In this paper, we present
\system{}, a network telemetry system that uses a uniform query interface to
drive the joint collection and analysis of network traffic.  \system{} takes
advantage of two emerging technologies---streaming analytics platforms and
programmable network devices---to facilitate joint collection and analysis.
\system{} allows operators to more directly express network traffic analysis
tasks in terms of a high-level language. The underlying runtime  \system{}
partitions each query into a portion that runs in the switch and another that
runs on the streaming analytics platform, iteratively refines the query to
efficiently capture only the traffic that pertains to the operator's query, and
exploits sketches to reduce state in switches in exchange for more approximate results.
Through an evaluation of a prototype implementation, we demonstrate that
\system{} can support a wide range of network telemetry tasks with less state in
the network, and lower data rates to streaming analytics systems,
than current approaches can achieve.
\end{abstract}

\section{Introduction}

Network operators routinely perform a variety of measurement tasks,
such as diagnosing performance problems, detecting network attacks,
and performing traffic engineering.  These tasks require collecting
and analyzing measurement data---often in real time---through a
process called {\em network telemetry}~\cite{NT-ietf16}. Historically,
this process has involved a distinct separation between collection and
analysis of network measurements, leading to data that is often too
coarse or too fine-grained to support a particular query or task.  For
example, when analyzing the performance of streaming video traffic
across a backbone link or interconnection point, an operator typically
has to cope with coarse-grained flow records, as opposed to detailed,
packet-level information that could provide insight about delays and
loss. Similarly, detecting an intrusion or denial-of-service attack
may require analysis of packet payloads, which also may be difficult
to execute at high traffic rates.

The first problem with current approaches is that analysis ordinarily only
begins after the data have been collected. Typically, an operator has a
question pertaining to some operational task and must make do with a
warehouse of packet-level or flow-level data that has already been collected.
The data is thus not well-suited to the query, and collection cannot adapt in
response to an operator's desire to refine a query. The second problem is that
the collection of the measurements themselves are constrained by the
(relatively limited) capabilities of current switch hardware, which generally
support static (and often coarse) data collection such as packet sampling
and simple counting.

In this paper, we argue that {\em queries about network traffic should
  drive both the collection and analysis of network measurements}.
Some existing systems~\cite{gigascope,kentik,deepfield} grapple with
the challenge of joint collection and analysis, but supporting a
general query interface efficiently and accurately on
networks of high-speed switches remains an unsolved problem.
Towards this goal, we present a packet-level telemetry system called
\system (Streaming Network Traffic Analysis).  \system exposes a query
interface with a familiar programming paradigm using dataflow
operators over the raw packet stream; the underlying runtime then
compiles the query to functions that operate on the switches, which in
turn pass a subset of traffic statistics to a scalable stream
processor. \system offers three new features:

\begin{itemize}
\itemsep=-1pt
\item {\em Uniform programming abstraction.} (Section~\ref{sec:applications}) Rather 
than relying on
  custom code for different kinds of measurement data, \system
  supports a wide range of queries with a single, familiar programming
  abstraction: dataflow operators over packet tuples, capitalizing on
  scalable stream processing architectures (\eg, Flink, Spark
  Streaming, Tigon)~\cite{flink,spark,tigon}.

\item {\em Query-driven data reduction with programmable switches.} (Section~\ref
{sec:collection})
  \system exploits the emergence of programmable network
  devices~\cite{rmt,p4-xilinx,barefoot-blog,p4-netronome} that can be programmed
  via a domain-specific language like P4~\cite{P4}. These devices
  support operations such as filtering, sampling, aggregation, and
  sketching at line rate to reduce the amount of data that the stream
  processor must handle.

\item {\em Coordinated data collection and analysis.} (Section~\ref{sec:coord})
\system
analyzes each
  query to coordinate the collection and analysis of the traffic. This
  creates significant gains in scalability, since the traffic that
  pertains to any given measurement task is generally a minute
  fraction of the overall traffic.
\end{itemize}

\noindent
Yet, simply juxtaposing a stream processing system with programmable
network devices is not, by itself, a viable solution. Coupling these
capabilities requires solving challenging design problems, such as:
determining which parts of the query should run on the switches and
which should run on the stream processor ({\em query partitioning}),
how to compute accurate estimates using limited-state ({\em sketches}),
and how and when to modify a query to zero-in on more fine-grained
subsets of the traffic ({\em iterative refinement}). These three
processes introduce tradeoffs between  the resources at switches
(\eg, state) and the stream processor (\eg, bandwidth and computation)
to execute the query and the latency to satisfy the query and the
accuracy of the result.  These tradeoffs depend on the workload,
such as the number of ongoing flows or the fraction of packets with
certain properties; thus, \system learns an efficient query
plan by solving an optimization problem that minimizes a weighted
objective function based on historical traffic patterns.

We use realistic network telemetry tasks over different real-world workloads,
to demonstrate that coordinated data collection and analysis help scale
query executions. Compared to state-of-the-art solutions that rely on the data 
plane for filtering and sampling only, \system reduces the load on the stream 
processor by more than a factor of four. Compared to solutions that exploit sketches 
to reduce state in the data plane without coordinating with stream processors, \system 
reduces the amount of state required by more than a factor of two.

\section{Current Network Telemetry Approaches}
\label{sec:background}

We review the state of the art in network telemetry.
Whereas \system{} uses queries to \emph{jointly} perform data collection and
analysis, existing network monitoring systems primarily tackle {\em either}
collection or analysis, with analysis typically occurring only after
collection. This section surveys the state of the art in
collection and analysis separately.

\subsection{Traffic Collection} 
\label{ssec:coll}
Network traffic collection and monitoring falls into two classes: packet-level
monitoring (sometimes referred to as ``deep packet inspection'') and flow-level
monitoring.

\smartparagraph{Packet monitoring} Packet-level monitoring can be performed
with software libraries such as {\tt libpcap}, or in hardware, using
devices such as the Eagle~10 or Endace capture
cards~\cite{endance}. Commonly, collection infrastructure is deployed
on a switch span port, which mirrors traffic going through the
switch. A device connected to the span port---typically a
server---captures and stores the mirrored traffic. The collection
infrastructure can be configured with filters that can specify
conditions for capturing traffic; configuration can also determine
whether complete packet payloads are captured, or simply an excerpt of
the packet, such as packet headers. Packet-level monitoring can
provide precise information for calculating statistics like the
instantaneous bitrate, packet loss, or round-trip latency experienced
by individual flows. Access to packet payloads can also be useful for
a variety of purposes, such as determining whether any given packet
carries a malicious payload.

Unfortunately, packet-level monitoring has significant drawbacks, due
to the high overhead of collection, storage, and analysis. The sheer
volume of network traffic makes it prohibitive to capture every
packet. Even if the infrastructure could capture every packet,
operators face daunting storage and analysis hurdles associated with
storing a complete log of all network traffic. As such, despite the
rich possibilities that packet-level traffic capture offer, many
networks do not deploy this type of infrastructure on a widespread
basis. For example, recent figures from a large access ISP have
indicated that deep-packet inspection capabilities are deployed for
less than 10\% of the network capacity. This sparse
deployment makes it essentially prohibitive to generally perform the
types of queries involving network performance or security that could
benefit from packet-level statistics.

\smartparagraph{Flow monitoring}
An alternative to packet-level monitoring is flow-level
monitoring---standardized in the Internet Engineering Task Force
(IETF) as IPFIX, and commonly referred to by the Cisco ``NetFlow''
moniker.  IPFIX permits each switch to collect flow-level statistics
that contain coarse-grained information such as the number of packets
and bytes for a particular flow (\eg, as defined by the source and
destination IP address, source and destination port, and protocol), as
well as the start and end time of the flow. This type of information
is often gathered in a ``sampled'' fashion: on average, one out of
every $n$ packets is tabulated in an IPFIX flow record; typical
sampling rates for an ISP backbone network can be in the 1,000 $< n <$
10,000 range, meaning that low-volume flows may often not be captured
at all.  Additionally, IPFIX records do not contain detailed
information about flows, such as packet loss rates or packet timings,
let alone packet payload information.

Both packet-level and flow-level monitoring systems can, of course, be
tailored to capture specific subsets of traffic. Packet-level monitoring can
be customized with filters that focus on specific subsets of traffic, and
flow-level monitoring can be tuned so that sampling rates are higher for
specific links of interest.  The advent of programmable data planes has also
enabled programmatic collection of individual data flows~\cite{bigtap,
univmon, opensketch, everflow, pathdump, narayana2016codesign, fb-p4}. Yet,
these tools either support limited sets of queries that can only operate over
fixed packet headers (\eg,~UnivMon~\cite{univmon} and OpenSketch~\cite{opensketch})
or require custom tools to analyze one specific type of data (\eg,
BigTap~\cite{bigtap} and PLT~\cite{fb-p4}) precluding any analysis that requires fusing
multiple data streams.

Furthermore, the level of flexibility that all of these systems offers is
limited, in the sense that (1)~in general, their configurations remain static;
(2)~decisions about capturing more fine-grained information are completely
decoupled from the queries or analysis, which occurs {\em post facto}.  In
short, because the monitoring process is decoupled from analysis, all of these
decisions must be made far in advance of analysis, often resulting in traffic
collection that is either too sparse or too voluminous.

\begin{table*}[t]
\begin{footnotesize}
\begin{center}
\begin{tabular} {| m{2.5 in} m{3.5 in} |}
\hline
{\bf Query} & {\bf Description} \\ \hline
\multicolumn{2}{|l|}{\textbf{Queries that process packet header fields}} \\ 
Heavy Hitter Detection~\cite{mina-snap, univmon, opensketch} & Identify 
flows consuming more than a threshold {\tt Th} of link capacity.\\
Superspreader Detection~\cite{mina-snap, univmon, opensketch}  & Identify 
hosts that contacts more than {\tt Th} unique destinations.\\
Port Scan Detection~\cite{minds} & Identify hosts
that send traffic over more that {\tt Th} destination ports.\\
SSH Brute Force Detection & Identify hosts receive similar-sized packets
from more than {\tt Th} unique senders.\\
%\multicolumn{2}{|l|}{}\\
\multicolumn{2}{|l|}{\textbf{Queries that process packet's payload}} \\ 
DNS TTL Change Tracking~\cite{mina-snap, exposure, chimera} & Identify hosts for which their domain's {TTL}
value changes more than {\tt Th} times.\\
IP-2-Domain Anomaly~\cite{mina-snap, exposure, chimera}  & Identify hosts that 
are shared by more {\tt Th} domains.\\
Domain-2-IP Anomaly~\cite{mina-snap, exposure, chimera}  & Identify 
domains that are advertised by more {\tt Th} hosts.\\
Sidejacking~\cite{mina-snap, exposure, chimera} & Identify  HTTP session 
cookies that are used by more than one hosts. \\
%\multicolumn{2}{|l|}{}\\
\multicolumn{2}{|l|}{\textbf{Queries that process packet's context fields}} \\ 
Loop Freedom Detection~\cite{pathdump, pathquery, everflow, fb-p4} & Detect forwarding loops.\\
Congested Link Detection~\cite{pathdump, everflow, fb-p4}  & Identify flows that traverse congested links.\\
Silent Packet Drop/ Blackhole Detection~\cite{pathdump, everflow,
  fb-p4}  & Identify switches that drop packets.\\
%JEN: confusing.  what is a "conformance policy"?
Path Conformance Detection~\cite{pathdump}  & Identify flows that violate path constraints, 
such as maximum path length.\\
\hline
\end{tabular}
\end{center}
\end{footnotesize}
\caption{Queries that we have written using \system's
packet-as-tuples abstraction.}
\label{tab:apps} 
\end{table*}

\subsection{Traffic Analysis}
\label{ssec:analysis}

Given the ability to perform packet or flow monitoring on network traffic,
network operators can use systems such as Deepfield~\cite{deepfield}, 
Kentic~\cite{kentik}, or Velocidata~\cite{velocidata} to
support network analysis in support of network performance or security.  For
example, Deepfield Singularity performs joint analysis of packet captures and
IPFIX records to help network operators understand questions such as the
relationship between traffic overload and application performance, as well as
detect network attacks such as distributed denial of service attacks.  For
example, analyzing the average bitrate of Netflix streams traversing the
network requires: (1)~capturing the DNS queries (and responses) for DNS
domains corresponding to Netflix streams; (2)~joining the resulting IP
addresses to the corresponding traffic data (\eg, either IPFIX or packet
capture) that can provide information about the rates that individual flows
are seeing. Another example might be the detection of a DNS reflection and
amplification attack, which involve compromised hosts sending large volumes of
DNS queries with the spoofed source IP address of the victim. Detecting such
attacks often involves detecting an abnormally high number of DNS queries from
an IP address (in this case, the IP address of the victim), typically for DNS
query types that elicit large responses (\eg, TXT, RRSIG); alternatively, one
could look for an abnormally large number of such responses destined for a
given IP address.

Although existing technologies developed by Deepfield and Kentik support
certain aspects of this type of analysis, they do not use the query itself to
drive collection of the traffic data, which often results in collecting,
storing and analyzing large volumes of data that do not pertain to the
specific queries. Specifically, these analysis tools rely on {\em separate}
collection of DNS data (with packet monitoring) and traffic utilization
information (\eg, with IPFIX), which the analysis tools subsequently joint
{\em post hoc}. This approach to analysis also requires capturing a large
amount of traffic that is not relevant to the analysis, which increases the
overhead of the analysis, both in terms of the volume of data and the
computation time. Furthermore, because IPFIX data is often based on sampled
traffic traces with high sampling rates, many DNS queries and responses will
not be captured in the IPFIX data at all, severely compromising accuracy.
Finally, the {\em post hoc} nature of existing analysis tools also precludes
real-time detection, since all data is collected and warehoused for subsequent
joint analysis.

%To confirm the attack, the network operator may look for evidence that
%the attack involves \emph{amplification}, where a small DNS request
%triggers a large DNS response.  To generate large responses, the
%adversary has the compromised hosts send DNS ``ANY'' requests that
%cause the DNS server to send a response where each resource record has
%a large (\eg, 1024-bit) signature~\cite{rossow}.  The network operator
%would like to check that these suspected victims are seeing DNS
%responses with such ``RRSIG'' resource records.  However, Netflow does
%not provide such fine-grained, application-layer information. So, the
%network operator would need to rely on packet monitoring.  However,
%the packet monitor may not be able to filter on the DNS resource
%record type, either, requiring the network operator to configure the
%packet monitor to capture all DNS response traffic sent to the
%suspected victims (or sample at some acceptable rate), and then check
%the DNS resource record type of these packets to settle on a final set
%of victim destinations.

\section{Uniform Programming Abstraction}
\label{sec:applications}

In this section, we first introduce the programming
abstraction for expressing queries, which is based on extensible packet-as-tuple
abstractions. Then, we show how applying
dataflow operators over these tuples can support a wide range of
network telemetry applications.

\subsection{Extensible Packet-tuple Abstraction}
\label{ssec:prog-abs}
%%%
%%% Packet tuple
%%%

Network telemetry involves answering questions about the packets
flowing through a network.  Packets not only carry information about
the header fields and payload, but also they can sometimes carry the
information about the state of the underlying network, as in in-band network
telemetry (INT)~\cite{INT}. Allowing network operators to express their queries
directly over packet tuples, where each tuple captures the properties of the
packet and its experience in the network, enables network operators to express
queries for a wider range of network telemetry tasks.  Existing systems such as
Everflow~\cite{everflow} and Pathdump~\cite{pathdump} also operate at
packet-level granularity, but only over a limited set of fields.

\system presents a simple abstraction where queries operate over all 
packets, at every location in their journeys through the network. Fields in the
packet tuples can include:

\begin{itemize}
\item Packet contents, including header fields in different protocol
  layers such as Ethernet (\eg, {\tt srcMac}), IP (\eg, {\tt srcIP},
  {\tt dstIP}, and {\tt proto}), UDP/TCP (\eg, {\tt srcPort}), and
  application (\eg, {\tt dns.ttl}), or even payload ({\tt
    payload})

\item Packet size in bytes ({\tt size})

\item Location ({\tt locationID}), which identifies a specific queue
on a specific switch in the network
%JEN: readers will ask about clock sync.
%JEN: omitting tin, tout, and qSize as we never talk about them later
%       and don't really support it yet.  good to avoid raising questions.
%\item The times the packet enters and leaves the queue ({\tt tin}
%and {\tt tout}), and the size of the queue when the packet arrives
%({\tt qSize})
\end{itemize}
\noindent
%Previous systems, like Everflow~\cite{everflow} and
%Pathdump~\cite{pathdump}, also operate at the packet-level granularity,
%but over a limited set of fields.
%%%
%%% User-defined parser
%%%

\noindent
To allow the operator to extend the tuple abstraction (\eg, by adding
support for protocol-specific tuple values), \system's parser specifies
how to extract the packet-level information necessary to support a
specific set of queries.  For example, a query on DNS traffic may need
to extract information from DNS messages (\eg, {\tt dns.rr.type}),
whereas some security applications may analyze the packet payload
({\tt payload}) as a string. The ability to parse traffic at line rate often depends
on the capabilities
of the respective data-plane targets.
When compiling
queries, \system aims to perform parsing in the underlying switches
whenever possible, directing certain packets to the stream processing system
for further processing only when necessary. \system is designed to easily incorporate
standard parsers for common protocols.

%%JEN: seems too implementation specific
%As a default, we support extraction of all the header
%fields supported by the data plane targets, and also payload fields at the
%stream processor using Scapy's schema~\cite{scapy-fields}. Note that even though
%the programming abstraction supports arbitrary fields in the packet tuple,
%parsing certain fields and the extent of a packet's available context, depends
%heavily on the capabilities of the underlying targets. 

%%JEN: not sure how to think about qSize, as it is awkward to have
%%something in the language that cannot necessarily be supported by (i) the
%% data-plane target or (ii) sending the packet to the stream processor.
%For example, not all
%targets can extract the {\tt qSize} field from the packet. Thus, queries over 
%{\tt qSize} field can only be supported over targets that can extract this field
%from the data plane.
 
%Streaming analytics platforms commonly operate on streams of tuples. In general,
%tuples can represent arbitrary data; conveniently, packets can be represented in
%tuple form. 

\begin{figure}[t]
      \begin{lstlisting}[language=Python,basicstyle=\footnotesize, 
caption= Detect potential victims of DNS reflection attacks by counting distinct sources.
%\texttt{p} denotes a packet tuple.
,
captionpos=b, label=asymm-query, captionpos=b,
basicstyle=\footnotesize, 
numbers=left,xleftmargin=2em,frame=single,framexleftmargin=2.0em]
pVictimIPs(t) = pktStream(W)
    .filter(p => p.srcPort == 53)
    .map(p => (p.dstIP, p.srcIP))
    .distinct(key=(dstIP, srcIP))
    .map((dstIP, srcIP) => (dstIP, 1))
    .reduce(key=(dstIP), func=sum)
    .filter((dstIP, count) => count > Th1)
    .map((dstIP, count) => dstIP)
\end{lstlisting} 
\end{figure}

\subsection{Expressive Dataflow Operators}
\label{ssec:dataflow}

Network telemetry applications often require collecting aggregate
statistics over a subset of traffic and joining the result of one
analysis with another. 
%\mcnote{Would express be a better word here? I feel we
% don't highlight enough that we consider a high-level declarative interface for specifying monitoring applications} 
Most of these tasks can be expressed as declarative queries composing
dataflow operators like {\tt map}, {\tt reduce}, and {\tt join} over a
stream of packet tuples. Unlike existing solutions where telemetry tasks are
tightly coupled with the choice of collection tool, our programming abstraction 
hides the details of how \system performs query execution, {\em where} each
query operator runs, and {\em how} the underlying targets perform
operations.  Thus, the same queries can be applied over different
choices of streaming or data-plane targets---ensuring that the
telemetry system is flexible and easy to maintain.

%\mcnote{It might help to mention explicit that we build upon previous work here, like DryadLINQ, Spark, etc. as we were saying in the NSDI submission. We don't need to calim any novelty about dataflow so it'd help clarify what we are assuming are building blocks that already exist.}
\system's query interface is inspired by dataflow frameworks like Spark~\cite
{spark}; we currently support {\tt map}, {\tt reduce}, {\tt distinct}, {\tt
  filter}, {\tt sample}, and {\tt join} operations over the stream of
packet tuples.  Stateful operators like {\tt distinct} and {\tt reduce} are
applied on the stream over a rolling time window (of $W$ s) specified in the query;
every $W$ seconds, the values of those operators are evaluated and reset. Queries can also express a maximum acceptable delay (in seconds) for detecting statistics of interest, as well as an error tolerance for the answers. Tolerating bounded latency and error gives \system flexibility to
answer queries efficiently with limited resources, as discussed in Section~\ref{ssec:sketches}.

\begin{figure}[t]
\begin{lstlisting}[language=query2,basicstyle=\footnotesize,  escapechar=@,
basicstyle=\footnotesize, numbers=left, label=payload-query, xleftmargin=2em,
frame=single,framexleftmargin=2.0em, captionpos=b, 
caption=Confirm reflection attack victims based on DNS header fields. \texttt{pVictimIPs(t-W)} refers to the stream of Query~\ref{asymm-query} at the previous interval $t-W$.]
victimIPs(t) = pktStream(W)
  .filter(p => p.srcPort == 53)
  .filter((p => 
      pVictimIPs(t-W).contains(p.dstIP))
  .filter(p => p.dns.rr.type == 46)
  .map(p => (dstIP, 1))
  .reduce(key=(dstIP), func=sum)
  .filter((dstIP, count) => count > Th2)
  .map((dstIP, count) => dstIP)
\end{lstlisting}
\end{figure}
  %.filter(p => p.proto == 17)
%\mcnote{Why distinct has no parameter? Shouldn't it be the field that is tracked for distinct values?}
%\mcnote{A similar issue is with the ``sum'' function passed as a parameter. Not too clear what is being summed.}

Applying dataflow operators over packet tuples makes it relatively
easy to express telemetry queries, such as detecting the onset of the
DNS-based reflection attacks discussed in Section~\ref{ssec:analysis}.
One way to detect DNS reflection attacks is to identify destinations
({\tt dstIP}) receiving DNS responses ({\tt srcPort} of 53) from a
large number (more than {\tt Th1}) of unique sources ({\tt
  srcIP})~\cite{polychroniou2014track}, as shown in
Query~\ref{asymm-query}. Then, Query~\ref{payload-query} operates over
DNS header fields to confirm the presence of reflection attacks.
After winnowing out the destinations that do not satisfy
Query~\ref{asymm-query} during the previous window interval (line
$3$), Query~\ref{payload-query} looks only at DNS response messages of
type RRSIG ({\tt dns.rr.type} of 46)~\cite{ietf-rrsig}, commonly
associated with amplification attacks~\cite{rossow}.  The query then
keeps a {\tt count} for each {\tt dstIP} that receives such response
messages, returning the set of victims that exceed some threshold
({\tt Th2}).  Writing these high-level queries is much simpler than
performing custom analysis of sampled Netflow data (for
Query~\ref{asymm-query}), and configuring a packet monitor and
analyzing the results (for Query~\ref{payload-query}).
%\mcnote{This seems over claimed. If I used OpenSOC or Tigon it would not be that different from the above?}

Table~\ref{tab:apps} summarizes some of the queries that we have written using
\system. These examples, drawn from the
existing literature, illustrate the expressiveness of the packet-tuple
abstraction combined with dataflow programming to support a wide range of
telemetry applications.

%For path related
%queries, we collect (analyse) the data from the edge switches, similar to existing works
%such as Pathdump~\cite{pathdump}\footnote{Unlike Pathdump~\cite{pathdump}, 
%that executes queries at the edge server, \system executes queries over the edge 
%switches and stream processor.} and PLT~\cite{fb-p4}. 
%We also assume that the intermediate switches update the {\tt path} metadata for 
%each packet---leveraging the data plane targets that support in-network telemetry 
%(INT)~\cite{fb-p4, INT}. 

\section{Query-Driven Data Reduction}
\label{sec:collection}

We now describe the capabilities of PISA
(Protocol-Independent Switch Architecture) targets and explain
how \system supports dataflow operations on these targets.

\subsection{Protocol-Independent Switch Architecture (PISA)}
\label{ssec:pisa}

Conventional network devices do not support custom packet processing or
state management.  In contrast, recently introduced PISA
targets~\cite{rmt,p4-xilinx,barefoot-blog,p4-netronome} provide
features that can support dataflow operations directly in the data
plane, implementable in the P4 language~\cite{P4}.

\smartparagraph{Programmable parsing.}  PISA targets allow for the
specification of new header formats for parsing packets.  Programmable parsing
enables both the extraction of desired header fields for answering arbitrary
queries and the definition of application-specific header formats for sending
streams of tuples to a stream processor.

\smartparagraph{State in packets and registers.} 
PISA targets have registers that support simple stateful
computations, as well as match/action tables with a byte and packet
counter for each rule.  These capabilities can allow queries to
accumulate statistics across a sequence of packets (\eg, a sum in a
{\tt reduce} operation).  In addition, PISA targets can place state
in custom metadata that is carried along with a packet through the
packet-processing pipeline or on to the next switch, enabling queries
to perform more complex operations across multiple stages.

\smartparagraph{Customizable hash functions.} 
PISA targets support hash functions over a flexible set of fields,
\eg, to access a specific register in a register array.  These hash
functions are useful for implementing operators like {\tt reduce} and
{\tt distinct} that maintain state that depends on combinations of
{\em query-specific} reduction keys.

\smartparagraph{Flexible match/action table pipelines.} PISA targets
support flexible match/action tables with programmable actions.  The
flexible matches on packet-header fields can support {\tt filter}
operations, and programmable actions enable the computations that
update state or affect the next stage of packet processing.

\subsection{Compiling Dataflow Operators to PISA Targets}
\label{ssec:choices}
Using these features, PISA targets can directly support many
dataflow operators. For a PISA switch, \system 
can implement the following operations.

\smartparagraph{{Map}, {filter}, and {sample} operations.}  Filter
operations, like line 2 of Query~\ref{asymm-query}, match on fields in
the packet's header. Naturally, this operation aligns with a
match-action table in the data plane; specifically, line 2 corresponds to a
table that matches on the {\tt proto} field and either permits the packet to
continue processing or ignores it. Map and sample operations can also be
executed using similar match-action tables each applying different actions over
matched packets. For example, in Query~\ref {asymm-query}, line 5 corresponds to
an action in a table that transforms the {\tt dstIP}, {\tt srcIP} pair into a
tuple of ({\tt dstIP}, $1$). Operations that require evaluating a predicate, as
in line 7 of Query~\ref{asymm-query}, are implemented with
%JEN: why are two stages needed for line 7, when only one of the two
% outputs of the predicate requires any action to be taken? also, it
% is a little confusing to say "Operations that require evaluating a
% predicate", as that statement is also true of line 2.
two match/action
tables: one for each possible evaluation of the predicate. For all of these
operations, the state required to execute them in the data plane consists of the
entries in the match/action tables.  The savings in processing realized at the
stream processor comes at the cost of maintaining this state in the data plane.

\smartparagraph{Distinct and reduce operations.}
The {\tt distinct} and {\tt reduce} operations are slightly different
from the previous operations because these operations require
maintaining state across sequences of packets.  In the case of {\tt
  distinct}, the state maintained is a single bit indicating whether
or not a particular key has already been observed; in the case of {\tt
  reduce}, it is the result of applying a function (\eg, {\tt sum})
over a particular (set of) key(s). We use hash tables (implemented as
arrays of registers) to maintain cross-packet state and metadata
fields for storing and updating the values from the hash tables. The
state required for executing these stateful operations in the data
plane can be quantified as the total number of registers used.

%\smartparagraph{Join operations.} Supporting {\tt join} operations in the data plane
%is difficult as it entails additional complexity of supporting build and probe
%semantics for implementing doubly pipelines {\tt join} operator~\cite
%{aqp-survey}. We currently execute {\tt join} operation as {\tt filter} where
%the output of the left query from the previous window interval is served as a
%filter for the right query. For example, in case of Query~\ref{payload-query},
%the output of Query~\ref{asymm-query}, \ie~anomalous {\tt dstIPs} are 
%applied as filter operation in line 4 of Query~\ref{payload-query}.

\smartparagraph{Limitations.}
PISA targets cannot support all of \system's dataflow operators directly in the
data plane. The set of supported parsing actions and available computational
capacity limit these targets' ability to support arbitrary dataflow operations.
For example, extracting a payload and performing arbitrary regular expression
matching over the payload is not currently supported in PISA targets. Also,
only simple computations, like {\tt add} and {\tt subtract}, or simple bit
manipulations (\eg, {\tt bit\_or}), can be applied over tuple fields
in the data plane. Thus, operations that require complex transformations over
tuple fields~\eg, a {\tt reduce} operation for entropy estimation which requires
a logarithmic transformation, cannot be supported in the data plane.

%\textbf{Relaxing accuracy with sketches}
\subsection{Bounding Data-Plane State With Sketches}
\label{ssec:sketches}
Maintaining the state required by these dataflow operations consumes
scarce memory and typically takes the form of a hash table. 
To avoid hash collisions, these hash tables are often both bloated and
sparse. Instead, we can employ probabilistic data structures that summarize
the relevant data in constant-space at the expense of a probabilistically-bounded error.
For example, Bloom filters~\cite {bloom-filters} are
compact data structures useful for set-membership testing.

In the case of the {\tt distinct} operator, we can use a Bloom filter
to test whether or not a given key is a member of the ``set of unique
keys''.  For example, in Query~\ref{asymm-query} at line 4, we use a
register array with $m$ rows storing only a single bit in each row.
For a single input key (\ie, {\tt dstIP} and {\tt srcIP} pair), $k$
hashed indices into this array are calculated.  The values at these
indices in the array determine whether or not a given key is already a
member of this set.  After checking set membership, these $k$ array entries
can be set to $1$ so future checks for the same input key are successful.

Similarly, when performing a {\tt reduce} with a sum, we can employ a
count-min sketch.  This sketch estimates the count associated with
each key by maintaining $k$ $m$-bit wide arrays, each indexed using
$k$ different hash functions.  For example, in
Query~\ref{asymm-query}, we use a count-min sketch at line 6 for
performing the reduction over the ({\tt dstIP}, $1$) tuples. For each
incoming packet, for the input key {\tt dstIP}, $k$ indices are
computed. The count values stored at each of these indices are
incremented by one and the minimum of the updated values is selected
as the estimated count. By carefully choosing the values of $k$ and
$m$, we can achieve significant size reduction with a provable
accuracy guarantee~\cite{cms}.

\subsection{Compiling Dataflow Queries to PISA Targets}
\label{ssec:query-compilation}
\smartparagraph{Compiling a Single Query.}
To compile a dataflow query to a PISA switch, we first configure the match/action
tables and hash tables required to execute the individual operators for the
query as discussed above. We then ensure that these operator-specific tables are
correctly applied in sequence. We take two goals into consideration while
compiling dataflow queries for PISA targets: (1)~The query processing pipelines
should not affect a packet's forwarding behavior; speficially, transformations over
a
packet's header should not be applied over the packet's actual header
fields. (2)~The mirroring overhead for each query should be minimal; specifically,
the switch
should only mirror packets that need to be reported to the streaming target. To
achieve the first goal, we create query-specific metadata fields---copying
information from the original header fields to these fields. To achieve the
second goal, we maintain one additional bit in metadata,~{\tt
report}, for each packet that specifies whether or not the result associated
with the packet should be reported to the streaming target.  For example in
Query~\ref{asymm-query}, if a given packet causes {\tt count} to increment without
crossing the threshold {\tt Th1}, there is no need to report the tuple to
the stream processor.  Queries are always
evaluated with respect to a {\em particular packet}, \ie, a packet is needed to
trigger the query processing pipeline in the data plane.

\smartparagraph{Compiling Multiple Queries.}
Most streaming systems execute multiple queries in parallel, where
each query operates logically over its own copy of the input
stream. Rather than actually replicating the packets, we execute
queries in the data plane \emph{sequentially}. To decide whether or
not to report a given query result to the stream processor, we take
the union of the {\tt report} bits from each query. This design choice
significantly reduces the mirroring overhead and requires only an
additional bit of state. When reporting data to the stream
processor, this approach requires that each single packet carries the
metadata fields for {\em all} reported queries in the
pipeline---making the task of decoding the packet at the stream
processor harder, since the stream processor must determine
which of the queries actually need to see the tuple.
%Additionally, our
%current implementation configures a separate match/action table for each
%operator. For some PISA targets, the number of match/action tables we can
%configure are limited. This problem can be addressed by sharing resources
%between multiple queries. We leave this exploration for future work. 

%\smartparagraph{Updating Query Executions.}
%At the end of every window interval, a given query's output is computed by the
%streaming target, which it sends to the runtime system. At the end of every
%window interval, a query's execution plan may proceed to the next refinement
%level. The runtime system processes the output and updates any {\tt filter}
%operations for partitioned queries~\ie~to support a finer refinement level. It
%then calls the {\tt compile} function to update the packet processing pipeline
%in the data plane. The data plane driver updates the respective match/action
%table entries as well as resets the values of stateful variables for each query.

\section{Coordinated Data Collection \& Analysis}
\label{sec:coord}

In this section, we describe how \system makes use of two scalability techniques:
(1)~query partitioning, and (2)~iterative refinement, to scale query execution. 
We then describe the problem of query planning, that determines {\em how}
and {\em where} each query should be executed. 

\subsection{Query Partitioning}
Section~\ref{sec:collection} describes how \system can execute dataflow queries 
in PISA targets. Not all operations for a query can be executed in 
the data plane. Either the switch cannot support them or the state required
exceeds what the switch can support. Thus, for each query, the runtime needs
to decide how to partition the query---executing portions of a query
directly in the data plane and the remainder in the stream processor. 

\begin{table}[t]
\begin{footnotesize}
\begin{tabular} {| m{.65 in} m{2.1in} |}
\hline
{\bf Driver API} & {\bf Description} \\ \hline
{\tt {\bf isSupported}(q)} & Returns whether the target can execute query {\tt q}.\\
{\tt {\bf getCost}(q,t)} & Returns normalized cost of executing query {\tt q} over 
the target, given training data {\tt t}. \\ 
{\tt {\bf compile}([q])} & Executes list of queries [{\tt q}] over the target.\\
\hline
\end{tabular}
\end{footnotesize}
\caption{Target-specific driver's API exposed to \system's runtime.}
\label{tab:driver-api} 
\end{table}

\smartparagraph{Partitioning Plans.}
Before deciding how to partition a query, \system's runtime must first identify
possible ways (\ie, {\em partitioning plans}), in which it can partition the
query,
and estimate the cost of each partitioning plan. 
For any dataflow query with $P$ operators, there are $P+1$ possible ways in 
which it can partitioning the query---each executing first $0,1,\cdots P$ operators 
in the data plane and the remainder in the stream processor. Here $0$ represents
the plan in which we execute all the dataflow operators in the stream processor
and $P$ represents plan where we execute all dataflow operators in the data plane. 
We use $\mathcal{P}=\{0,1,\cdots P\}$ to represent the set of partitioning plans 
for each query. 

The runtime relies on 
the API exposed by the data plane drivers, as shown in Table~\ref{tab:driver-api}, 
to determine what partitioning plans can be executed in the data plane and their 
cost. For example, if the runtime calls the {\tt isSupported} function on the 
partitioning plan, executing all operators for Query~\ref{asymm-query} for a 
PISA target, it returns true because all the constituent dataflow operators can 
be implemented in a PISA dataplane. Similarly, it calls the {\tt getCost} function 
to estimate normalized cost of executing a partitioned query in the data plane. 
If the output normalized cost $b$ and/or $n$ are greater than one, then the plan 
cannot be supported by the data plane and/or the stream processor. 

%Before compiling a given query, \system's runtime must first choose how to
%partition the query.  Partitioning a query saves network bandwidth and
%processing at the stream processor, but comes at the cost of maintaining state
%in the data plane.  In section~\ref{sec:learning}, we will explore how the
%runtime chooses optimal partitions in a broader query execution plan, but for
%now we statically describe how \system's runtime interacts with the data plane
%driver to partition queries.  \system offers modular data plane support through
%the driver API described in Table~\ref{tab:driver-api}.  As long as a given data
%plane target implements the API, \system's runtime can support partitioning
%queries to that target.  For example, if the runtime called the {\tt
%isSupported} function on Query~\ref{asymm-query} for a PISA target, it would
%return true because all the constituent dataflow operators can be implemented in
%a PISA dataplane. However, let us assume that only {\tt filter}, {\tt map}, and
%{\tt distinct} were supported operators in the data plane.  In this case, the
%runtime would sequentially generate subqueries of Query~\ref{asymm-query} to
%find the maximum set of operators that the data plane can support.  In our
%hypothetical scenario, \system's runtime would determine that only lines 2-5
%could be partitioned to the data plane and that lines 6-8 would have to be
%partitioned to the stream processor.

\subsection{Iterative Refinement}
For many queries, the traffic of interest
is typically only a small fraction of the total traffic. If we continuously 
collect and analyze all the incoming packets, we end up wasting compute
resources and maintaining state for traffic that is of no interest as far as a
given query is concerned.  We argue that coupling collection and analysis for network 
telemetry helps \system to selectively spend resources on only the relevant portions of 
the traffic.

To this end, we exploit the hierarchical structure
of some of the fields in the packet tuple---{\tt srcIP}, {\tt dstIP}, {\tt
dns.qname}---executing queries at coarser levels of refinement for these
fields, and iteratively zooming-in on traffic that satisfies the query at a coarser 
level. Executing queries at coarser levels requires less resources in the data
plane. For example, if we replace the field {\tt dstIP} with {\tt dstIP/8} for 
Query~\ref{asymm-query} (see line 3 in Query~\ref{query-transform}), then the 
number of unique pairs of {\tt dstIP/8}, and {\tt srcIP} drops significantly. 
Iterative refinement winnows out the
uninteresting traffic in each round which, in turn, allows us to dedicate
available compute resources to the portion of traffic that matters for the
query. This process saves state in the data plane but comes at the cost of the
additional detection delay incurred during the iteration.

\smartparagraph{Refinement Keys.}
To enable iterative refinement, the runtime needs to find fields in the packet tuples
that can be used for iterative refinement (the~{\em refinement keys}). They are 
determined by identifying the set of fields that (1)~are used as keys in stateful 
dataflow operations like {\tt reduce} and {\tt distinct} and (2)~have a hierarchical 
structure that allows us to replace them with coarser versions without
missing any traffic that satisfies the original query. For example, Query~\ref
{asymm-query} can use the field {\tt dstIP} as a key for iterative refinement. 

While it is possible to have more than one candidate field for iterative
refinement, it is also possible to not find any candidate for refinement for a
query. Consider the example of the query for detecting sidejacking attacks 
(see table~\ref{tab:apps}). This query cannot benefit from iterative refinement
because the two fields {\tt sessionID} and {\tt userAgent} that are used in the 
stateful operation do not have any hierarchical structure.

\smartparagraph{Refinement Levels.}
After identifying the refinement key(s), the runtime needs to enumerate the
levels for each field that can be used for iterative refinement. To generalize, 
each refinement key ($R$) consists of a set of levels $\{r_1 \dots r_n\}$ where $r_1$ 
is the coarsest level and $r_n$ is the finest; thus, $r_1 > r_n$ implies that $r_n$ 
is ``finer'' than $r_1$. The meaning of the $n^{th}$ refinement level is specific to 
each key; in the case of an IPv4 address, the $r_1$ refinement level would 
be a {\tt /1} mask applied to the address. In the case of a domain name, the $r_1$ 
refinement level would be the root domain ({\tt "."}) and the $r_n^{th}$ refinement 
level would be a fully qualified domain name.

\subsection{Query Planning}
%Query planning for processing stream data has been explored extensively in the 
%database and networking communities~\cite{carney2002monitoring, sullivan1996tribeca, 
%progme, gigascope, abadi2003aurora, abadi2005design, amann2014count}.  

\system's query planning performs cost-based optimization to determine {\em
where} and {\em how} to execute input queries.  Query planning for \system
primarily focuses on deciding how to combine different
techniques like query partitioning and iterative refinement to make best use of available 
resources. 
For a single query, there exist many possible combinations of partitioning
plans and refinemet levels and selecting poor combinations of these might
actually worsen system performance.  We call the specific sequence in
which we refine each query and partition the query for each refinement
level while configuring sketches in the data plane, a {\em query
plan}. 
%Each plan will require different amounts of resources from the software
%and the data plane targets and each plan will incur a different detection delay
%based on the number of refinement levels. 
In this section, we describe how the
runtime selects the best plan among all the possible candidates for each
query.

\subsubsection{Workload-Agnostic Query Planning}
\label{ssec:strawman}
Let us first consider a solution in which the runtime selects a uniform {\em
query plan} for all input queries, without taking into consideration the
workload that affects each specific query differently. Selecting such a query 
plan might not help scale query executions for two reasons.

First, different refinement levels will winnow
out traffic at varying degrees of effectiveness depending on the fields used as
keys for {\tt reduce} and {\tt distinct} operations, the threshold values for 
{\tt filter} operations, and the workload-specific subset of traffic over which
the query is applied. If a given query plan iterates over refinement levels that
do not effectively filter out the uninteresting traffic, then both detection
delay and state in the data plane are increased. 

Second, sketch-based data structure performance would also decrease
because their guarantees on accuracy rely on an estimate of the true count of
elements to be stored therein. Therefore in order to determine which query plan
strikes the right balance of costs given the system and query-specific
constraints, we need traffic data collected from the network. 

\begin{table}[t]
\begin{footnotesize}
\begin{tabular} {| m{.1 in} m{2.7in} |}
\hline
% & {\bf Queries} \\
$Q$ & Set of input queries expressed using \system's API\\
$W$ & Interval for windowed operations for each query\\
$D_{max}$ & Max. detection delay specified for each query\\
% & {\bf Training Data} \\
$T$ &  Training data collected for $M$ window intervals\\ 
% & {\bf System Constraints} \\
$B_{max}$ & Max. state (bits) data plane can support\\
$N_{max}$ & Max. tuples that the streaming target can process in $W$ seconds\\
\hline
\end{tabular}
\end{footnotesize}
\caption{Notation.}
\label{tab:input} 
\end{table}

\subsubsection{Workload-driven Query Planning}
\label{ssec:problem-form}

The runtime system's goal is to select the minimum-cost query plan, using
traffic data collected from the underlying network for comparing the cost of
various candidate plans. We observe that if we think of all the specific
combinations of refinement levels and partitioning plans as vertices in a graph,
then all possible paths from the coarsest to the finest refinement level become
the {\em candidate query plans}. Better yet, the best query plan becomes the
shortest-path with minimum cost in the graph of candidate query plans. 
%We,
%therefore, choose to frame our optimization 
%problem in graph-theoretic terms.  

\smartparagraph{Generating Query Plan Graph.} 
Candidate query plans are paths embedded in a {\em query plan graph} ($g$). 
%Before we can describe the structure of this graph, let us first enumerate
%other sets from which we will construct the vertices of this graph.
%For each query $q \in Q$, let the set of all possible consecutive window intervals,
%used for iterative drill down, be defined as $\mathcal{I} = \{1, 2, \cdots \}$.
The maximum number of window intervals is the minimum of $D_{max}$, the maximum
detection delay each query can tolerate, and the number of possible refinement
levels. All possible combinations of refinement levels and partitioning plans that are 
executed in consecutive window intervals 
%can then be expressed as three-tuples
%of the form $\{(r,p,i) \mid r \in \mathcal{R}, p \in \mathcal{P}, i \in \mathcal
%{I}\}$ and 
constitute the vertices in the query plan graph. The query plan graph
is constrained by two requirements: (1)~edges are permitted only from vertices
with coarser refinement levels to finer ones, and (2)~from vertices in the $i^
{th}$ window interval to vertices in the $(i+1)^{th}$ window interval. 
Two vertices are connected by an edge in this graph if and only if
they are direct neighbors in a candidate query plan. After adding all the edges
that satisfy the above requirements, we discard all the disconnected vertices to
obtain the final query plan graph $g = (V_g , E_g)$ for query $q$. We also
define a globally ``coarsest'' vertex ({\tt Src}) and a globally ``finest''
vertex ({\tt Tgt}) as the source and sink for every candidate refinement plan as
vertices in $V_g$.

\begin{figure}[t] \centering \includegraphics[scale=0.5]{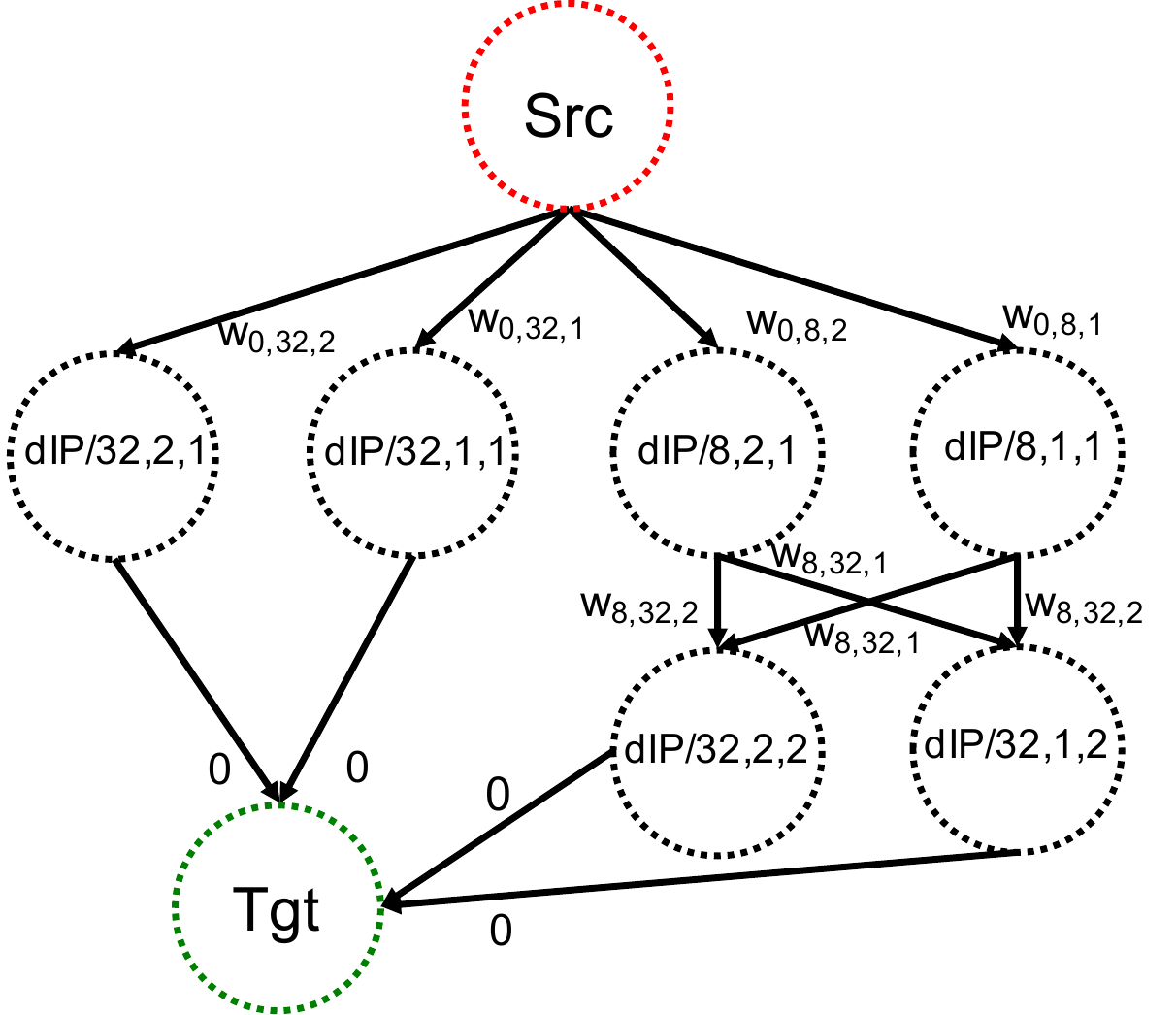} \caption{
Query plan graph for Query~\ref{asymm-query}. 
%Here refinement levels are 
%$\mathcal{R}=\{$({\tt dstIP/8}), ({\tt dstIP/32})$\}$, partitioning plans are 
%$\mathcal{P}=\{1,2\}$, and iteration rounds are $\mathcal{I}=\{1,2\}$. 
\label{fig-hgraph}}
\end{figure}

Figure~\ref{fig-hgraph} shows the query plan graph for 
Query~\ref{asymm-query}. It has {\tt dstIP} as the refinement key, 
and for ease of illustration has only two refinement (\{{\tt dstIP/8}, {\tt dstIP/32}\}) 
and partitioning (\{1,2\}) choices. In this example, there is no edge
between nodes {\tt  (dstIP/32},{\tt 1},{\tt 1}) to {\tt (dstIP/8},{\tt 1},{\tt2}) because 
we only allow zoom-in from a coarser to a finer refinement level in
successive iterations. Not shown are the disconnected nodes we discarded ({\tt
dstIP/8},{\tt 1},{\tt 2}) or ({\tt dstIP/8},{\tt 2},{\tt 2}) from the final
query plan graph shown in Figure~\ref {fig-hgraph}. Every path in this graph
from {\tt Src} to {\tt Tgt} represents a candidate query plan.

\smartparagraph{Updating Weights for Query Plan Graphs.} 
We will now describe how the runtime uses the traffic data to update the weight
for each edge in the resulting query plan graph $g =(V_g ,E_g)$. We compute a
new weighted graph for every window interval in the traffic data. If the
duration of data is $M$ window intervals long, then we compute $m \in [1,M]$
different weights for each edge and generate $M$ different weighted query plan
graphs. 

To compute the weights for each edge in the query plan graph,
the runtime needs to generate a transformed version of the original query that
is appropriate for the specific refinement level and partition plan of each node
in the graph. The runtime first maps the refinement keys to coarser values to
count statistics at coarser levels of granularity (see line 3 in Query~\ref{query-transform}). 
It must also update the thresholds at each refinement level. Selecting
the appropriate threshold at coarser levels is challenging. For example, in 
Query ~\ref{asymm-query}, the network operator specifies a threshold value on line 
7 for a specific {\tt dstIP} (also written as {\tt dIP/32}). At coarser
refinement levels, such as {\tt dstIP/8}, it would be inappropriate to apply the
same threshold as originally specified because the field in question is now an
aggregate count bucket and not a specific {\tt dstIP/32}. \system's runtime has
to select appropriate thresholds for each node in the graph such that no traffic
that satisfies the query at the finest level is missed.  

The runtime first runs the query at the finest 
refinement level over the training data to identify the portion of traffic that satisfies 
the original input query. It then backtracks to a coarser refinement level, runs the 
query again, but this time tracks the results for aggregate count buckets that contain 
the results at the next finer refinement level.  It then selects the minimum
count, of all count buckets, as the threshold value for this coarser refinement level.  
The runtime continues this process until it reaches the coarsest refinement
level thereby ensuring that at each refinement level, it selects a threshold
greater than or equal to the original threshold without missing any traffic that
satisfies the query.

\noindent
\begin{table}[t]
\begin{minipage}{\columnwidth}
\begin{lstlisting}[language=Python,basicstyle=\footnotesize, 
caption=Query transformations for iterative refinement,
captionpos=b, label=query-transform, captionpos=b,
basicstyle=\footnotesize, 
numbers=left,xleftmargin=2em,frame=single,framexleftmargin=2.0em]
Q1_8(t) = pktStream(W)
    .filter(p => p.proto==17)
    .map(dstIP --> dstIP/8)
    .map(p => (p.dstIP, p.srcIP))
    ... 
Q1_32(t) = pktStream(W)
    .filter(p => p.proto==17)
    .filter(p => (p.dstIP/8 in Q1_8(t-W)))
    .map(p => (p.dstIP, p.srcIP))
    ... 
\end{lstlisting} 
\end{minipage}
\end{table}

The runtime then joins queries for pair of
nodes---ensuring that the query at a finer level is only applied over traffic
that satisfies the query at the coarser level. It then provides the training
data and joined query for each edge as input to driver's {\tt getCost} (see
Table~\ref{tab:driver-api}). For each edge, this function returns the number of
packet tuples to be processed by the
stream processor and the amount of state required in the data plane. The target
applies an implementation-specific cost model to estimate the number of bits
required.
For PISA targets, it estimates the cost as number of bits required to execute
stateful operations like {\tt distinct} and {\tt reduce} in the data plane---estimating
sketch sizes in the process. To enable comparison between these two cost metrics, 
the function returns the values $n$ and $b$ normalized with respect to system
constraints. The costs for edge ($i,j$) depend only on how well
the traffic not-satisfying the given query was winnowed out at refinement level 
($r_{i}$) before getting executed at refinement level ($r_{j}$) with a
partitioning plan $p_{j}$. This implies that multiple edges will have the same
cost---reducing the complexity of computing the weights for query plan graphs. 
We define the weight for each edge ($i,j$) as a linear
combination of the costs: $w_{i,j} =\alpha n_{i,j}+(1-\alpha)b_{i,j}$.
$\alpha \in [0,1]$ is
a tunable parameter that assigns relative importance to the two cost metrics. 
%These weight computations result in a {\em weighted} query plan graph $g_w$
%where the cost of each path (i.e., candidate query plan) between
%the {\tt Src} node at the coarsest level and the {\tt Tgt} node at the finest
%refinement level is defined as the sum of the (normalized) weights of the edges
%on that path.

\subsubsection{Selecting the Best Query Plan}
The runtime 
%now has $M$ different weighted graphs. For each window 
%interval $m$, it 
can apply the {\em Dijkstra} algorithm over the weighted 
graph to find the plan with minimum cost. For each time 
interval, the edges in the graph will have different weights. 
Different time intervals are likely to have different minimum-cost query plans. 
Thus, the goal is to find the query plan that best represents the minimum-cost
query plan for {\em all} of the training intervals. We select the
candidate query plan ($p$) that minimizes the root-mean square error, 
$RMSE(p)=\sqrt{\frac{1}{n}\sum_{i=1}^{m}{(cost(p)-cost(p_{i})})^2}$.
Here, $p_m$ represents the minimum-cost query plan for interval $m$ and the
cost for each plan is defined as: $cost(p) = \sum_{e \in E_g} \alpha n_{e}+
(1-\alpha)b_{e} $
for all edges $e$ in plan $p$. 
%
%To ensure that the query plan selection is not biased due to noise in the data, 
%we rely on {\em K-fold cross-validation}. It is a common practice when solving 
%(supervised) machine learning problems; where we `hold out" parts of the
%available data as a validation set to learn the query plan with minimum error
%across all data points. We also periodically collect data from the network to
%update plans for each query. 

\smartparagraph{Tuning $\alpha$.}
The tunable parameter $\alpha$ assigns relative 
importance to the normalized $N$ and $B$ costs where $N = \sum_{e \in E_g} n_e$
and $B = \sum_{e \in E_g} b_e$. As we increase
$\alpha$,
we trade off higher cost $N$ for lower cost $B$. For the given system
constraints, $N_{max}$ and $B_{max}$, each value of $\alpha$ can result in
four possible states depending on whether cost $N$ is greater than $N_
{max}$ and/or cost $B$ is greater than $B_{max}$. To tune $\alpha$, 
we start with $\alpha = 0.5$ (equal weight to normalized $N$ and $B$ costs) 
and perform a binary search with the goal of finding an $\alpha$-value where the 
weighted cost of the query plan across all $M$ intervals
is minimum while ensuring that the constraints $N \le N_{max}$ and $B \le B_{max}$ 
are satisfied for each candidate $\alpha$. Note that no feasible solution exists if any
$N > N_{max}$ and $B > B_{max}$.

\smartparagraph{Selecting Query Plans for Multiple Queries.} 
Multiple queries compete for limited resources, specifically, the amount of state
in the
data plane and processing power at the streaming streaming targets. The
problem of finding the right query plans for all queries that require minimum
resources and also satisfy the system constraints, can be mapped to the {\em bin
packing problem}~\cite{bin-pack}.
%, and is thus an NP-hard problem. 
We designed a 
simple algorithm for multiple queries that tunes a single value of
$\alpha$ for all queries together, which is a simple extension
of our algorithm for the single query case. It performs a binary search
exploring the value of
$\alpha$ that minimizes the total weighted cost and applies the constraints over the
total number of bits and tuples for all the queries. 
%Clearly, this algorithm
%will not find the optimal solution where different values of $\alpha$ for
%different queries might minimize the total weighted cost. 
%
%We use this approach as an illustration, but other more complex (supervised) machine
%learning 
Other algorithms can readily be substituted to learn the best query plan.
% and tune 
%$\alpha$, using the training data and system constraints. We leave that exploration
%for future works. 

%In our current solution, we configure independent packet processing pipelines 
%in the data plane---maintaining separate hash tables for each query. However, it 
%is possible to share some match/action tables and registers across different operators of the input
%queries---making better use of available resources. The problem of designing
%such packet processing pipelines and learning  the right execution plans for
%such a setup is exciting, but is beyond the scope of this work. We leave this as
%problem for future work. 

\section{Evaluation} 
\label{sec:eval}

In this section, we evaluate \system using realistic network telemetry tasks over
different real-world workloads. 
%We also run a number of micro-benchmarking 
%experiments to quantify the various performance tradeoffs associated with
%our design choices for \system.

\begin{figure}[t!]
\centering
\includegraphics[scale=0.75]{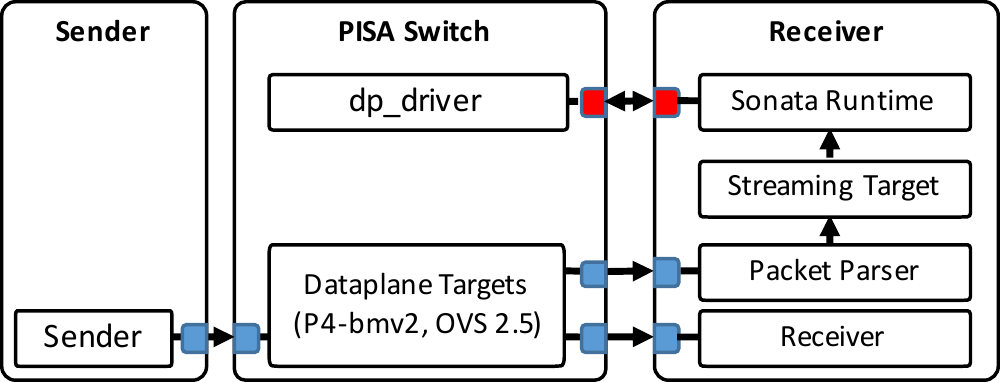}
\caption{Testbed used for evaluation. Blue interfaces are used for sending data packets
and the red ones are used for sending management traffic.}
\label{fig:setup} 
\end{figure}

\subsection{Setup}

\begin{table}
\begin{footnotesize}
\begin{center}
%\begin{tabular} {|l|p{0.75 in}|p{0.75 in}|p{0.75 in}|p{1 in}|p{1 in}|p{1 in}|}
\begin{tabular} {|ccm{1.7 in}|}
%& \multicolumn{3}{c|}{\textbf{Partitioning}} & \multicolumn{2}{c|}{\textbf{Refinement}}  &\\ 
\hline
{\bf Network} & {\bf Tool} & {\bf Description} \\ 
\hline
Large IXP &  IPFIX & Sample (1 in 10K packets), 3 Tbps peak rate,  August 2015.\\ 
Large ISP & NetFlow & Unsampled, 10 Gbps backbone link, January 2016.\\ 
Campus & Pcap & Unsampled, 150 Kbps DNS traffic, February 2017.\\ \hline
\end{tabular}
\end{center}
\end{footnotesize}
\caption{Real-world traffic data traces}
\label{tab:dataset} 
\end{table}

\begin{table}
\begin{footnotesize}
\begin{center}
%\begin{tabular} {|l|p{0.75 in}|p{0.75 in}|p{0.75 in}|p{1 in}|p{1 in}|p{1 in}|}
\begin{tabular} {|c|cc|m{1.4 in}|}
%& \multicolumn{3}{c|}{\textbf{Partitioning}} & \multicolumn{2}{c|}{\textbf{Refinement}}  &\\ 
\hline
{\bf  Queries } & \multicolumn{2}{c|}{\bf Lines of Code} &  {\bf Description}\\ 
& \system & P4 & \\ \hline
DDoS-UDP & 8 & 383 & Detecting traffic anomaly over UDP traffic\\
SSpreader & 7 & 333 & Superspreader detection query\\
PortScan & 7 & 320 & Port scan detection query \\
DDoS-DNS & 8 & 383 & Detecting traffic anomaly over DNS traffic, \ie~Query~\ref{asymm-query}\\
Reflection Attack & 16 & 402 & Detecting reflection attacks using DNS 
headers, \ie~Query~\ref{payload-query}\\\hline
\end{tabular}
\end{center}
\end{footnotesize}
\caption{Queries for telemetry tasks}
\label{tab:queries} 
\end{table}

\smartparagraph{Workloads.}
To quantify \system's performance, we use actual traffic traces collected from 
three different networking environments as shown in Table~\ref{tab:dataset}.
The first trace consists of sampled, flow-level statistics that was collected at
a large IXP using the IPFIX tool. The second trace is unsampled flow-level data
collected from a large ISP's backbone link connecting Seattle and Chicago. We
selected two hours of traffic from both of these two traces and used them
to evaluate the capability of \system's runtime system to learn
workload-driven query plans for scalability. The third
trace consists of unsampled DNS requests and responses that we collected from a
campus network. We use one minute of this low-volume packet trace to demonstrate
\system's end-to-end performance with software switches as data plane targets.

\begin{figure*}[t!] 
\centering
\begin{subfigure}[b]{.33\linewidth}
\includegraphics[width=\linewidth]{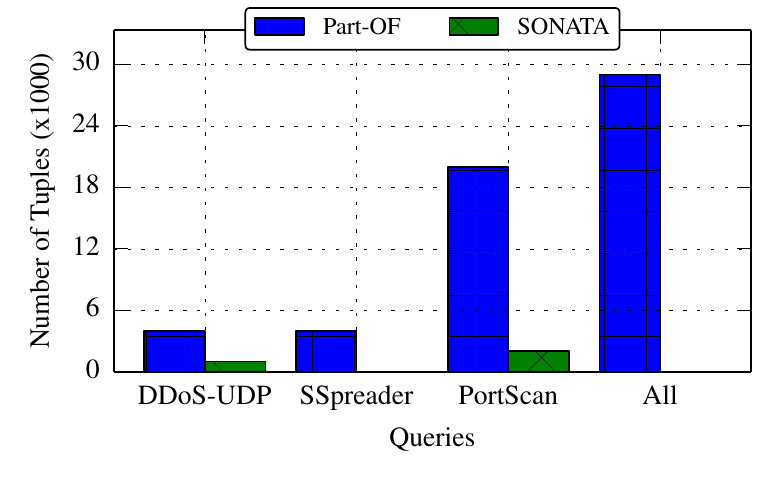}
\caption{Tuples at stream processor} 
\label{fig:tuples}
\end{subfigure}
\begin{subfigure}[b]{.33\linewidth}
\includegraphics[width=\linewidth]{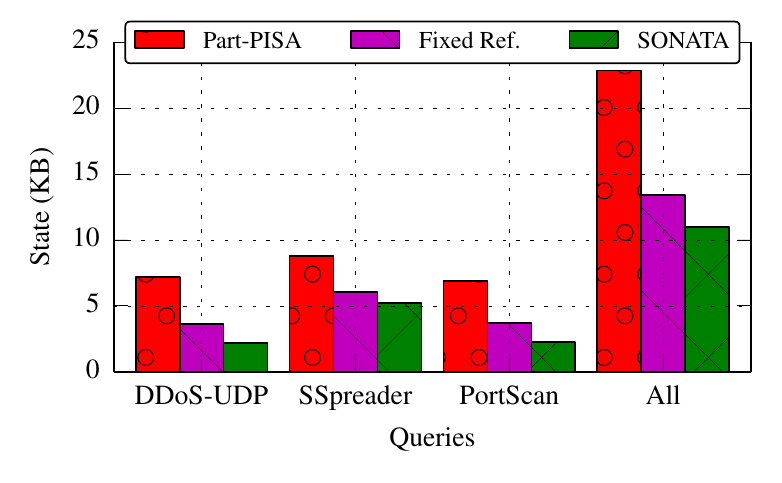}
\caption{State in data plane} 
\label{fig:bits}
\end{subfigure}
\begin{subfigure}[b]{.33\linewidth}
\includegraphics[width=\linewidth]{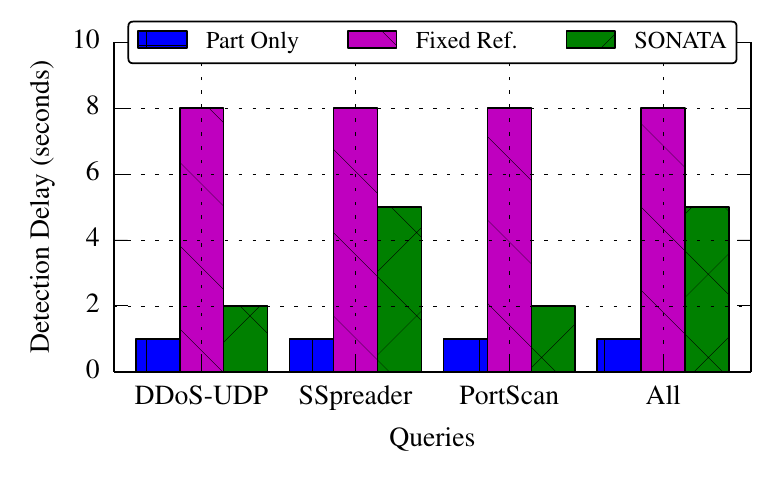}
\caption{Detection Delay} 
\label{fig:delay}
\end{subfigure}
\caption{Performance gains: (a)~number of tuples processed by the
stream processor, (b)~state required in the data plane, and (c)~time to
detect events of interests.}
\label{fig:perf-gains} 
\end{figure*}

\smartparagraph{Monitoring Queries.}
To compare and contrast \system's performance with different state-of-the-art
network telemetry solutions under realistic workloads, we primarily focus on 
queries that operate over packets-header fields. Applying these queries to the
available flow-level traces quantifies \system's performance under realistic
workloads.  These queries can also contrast \system's performance against
solutions that execute the entire query over the streaming or the data plane
targets. Table~\ref{tab:queries} lists the queries we use for our evaluation.
Unless specified otherwise, we set the threshold values for each query as the
$99.9^{th}$ percentile of the respective counts; window interval as one second;
and sketch accuracy as $99$\%. For all of these queries, the refinement keys are
IP addresses.  We consider eight refinement levels (\ie, \{{\tt /4}, {\tt /8},  
$\cdots$,{\tt  /32}\}) for iterative refinement. 

Table~\ref{tab:queries} also shows the lines of code for expressing these queries
using \system's query interface. We compare these numbers with the lines of code
required to configure match/action tables in the PISA targets to execute the
same queries. These results show that \system makes it easier for
network operators to express their queries without worrying about configuring
the low-level data plane targets. 

\smartparagraph{Prototype.}
Our prototype is implemented in {Python} (around 9,000 lines of code). We use 
the P4 Behavioral Model (bmv2)~\cite{bmv2}, and Open vSwitch 2.5
as the data plane targets, and use Spark 1.6.1~\cite{spark} as the stream
processor. We use scapy~\cite{scapy} for parsing the query-specific header
fields embedded in packet's header coming out from a switch's span port. 

\smartparagraph{Testbed.}
Our testbed has three physical servers, each with Xeon
4-core 3.70 GHz CPU, three 10 GbE, and one 1GbE cards. We use the 1G interfaces
for communication between different components (\ie, data plane driver and
runtime) and the other ports for sending data. Figure~\ref{fig:setup} shows how
we configured these machines. Note, in practice, PISA targets are expected to
support traffic rates up to 6 Tbps~\cite{tofino}. Unfortunately, this hardware
is currently still in a pre-release phase and not yet available to us. The
current P4 behavioral model, however, can only process a few hundred packets per
second. This limits our ability to perform end-to-end performance evaluation
using high volume workloads. 
%For this 
%reason, we only evaluate the performance of our bmv2-based prototype using our low-volume 
%campus network data (trace 3) as workload. 

\smartparagraph{Comparison.}
We compare \system's performance against four categories of state-of-the-art
network telemetry solutions. First, {\bf Stream-Only} represents solutions like 
OpenSOC~\cite{opensoc} that collect all raw packets for analysis. Second, {\bf Part-OF}
represents solutions like EverFlow~\cite{everflow} that can perform
limited data collection in the dataplane before analysis (\eg, partitioning {\tt
filter} and {\tt sample} operations). Third, {\bf Part-PISA} represents solutions such as
OpenSketch~\cite{opensketch} that can use PISA targetes for executing as many
dataflow operations as possible in the data plane. Fourth, {\bf
Fixed-Refinement} represents the class of solutions that apply a
workload-agnostic static refinement plan (see Section~\ref{ssec:strawman}) for
all queries in order to reduce the state required to execute dataflow operations
in the data plane.

\subsection{Scaling Query Executions}
We now demonstrate that using workload-driven query plans help scale
query executions. Compared to
configurations that rely on the data plane for filtering and sampling only, \system
reduces the load on the stream processor by more than
a factor of four.
Compared to approaches that exploit sketches to reduce sate in the data plane without
coordinating with stream processors, \system reduces the amount of
state required by more than a factor of two.

%To demonstrate that \system delivers on its promise to scale query executions,
%%Next we quantify how using learned query plans reduces the number
%%of tuples that have to be processed by the stream processor and the amount of state 
%%that is required by PISA targetes as our data plane targets. 
We apply both the large IXP and ISP workloads on the {\em DDoS-UDP}, {\em SSpreader}, 
and  {\em PortScan} queries. 
For each query, we configure the system constraints such that \system will
be able to partition at least one stateful operation to the data plane. For example, in the 
case of {\em DDoS-UDP} query , setting $B_{max}=2.5$ KB ensures
that all the incoming packets are not forwarded to the stream processor. 
For each experiment, we partition the workload data into two parts, using the first
twenty window intervals as training data for selecting the query plan, and the remaining
windows as test data. We report median values across all the test data points. 
Figure~\ref{fig:perf-gains} shows how \system scales the execution of the three
queries when they run independently and concurrently.

\smartparagraph{Number of Packet Tuples.}
Any gains compared to mirror-all-traffic solutions, \ie~{\em Stream-Only} are obvious.
Thus, in Figure~\ref{fig:tuples}, we show the number of packet tuples processed by {\em Part-OF}
and \system. We observe that compared to {\em Part-OF} solutions that can only execute
{\tt filter} and {\tt sample} operators to the data plane, \system's ability to execute stateful operations 
significantly reduces the load on the stream processor. Assuming the context of the scalability 
study in~\cite{osoc-scale}, which shows scalability results of the stream processor up to 1M packet 
tuples/s, by simply extrapolating the results for {\em DDoS-UDP} in this figure, we can expect that 
compared to {\em Part-OF} solutions that can support around 250 such queries, \system 
can support up to 1000 such queries.

\smartparagraph{State in the Data Plane.}
Figure~\ref{fig:bits} compares the amount of state required to execute a subset of the
query in the data plane by \system against {\em Part-PISA}, and {\em Fixed-Refinement}.
These solutions either do not use iterative refinement ({\em Part-PISA}) or use a fixed 
refinement plan for all queries ({\em Fixed-Refinement}). We can see that \system's ability 
to select workload-driven query plans ensures that it requires minimal state in the data plane.
%\mcnote{Next sentence seems incomplete: Given ..., then what?}
Given that in practice, 
PISA targets can easily support 4 MB of state in the data plane for monitoring 
applications~\cite{narayana2016codesign}, by simply extrapolating the results for 
{\em DDoS-UDP} query in this experiment, we can expect \system can support up to
1600 such queries compared to {\em Part-PISA} solutions which can only support 800. 

\smartparagraph{Detection Delay.}
Figure~\ref{fig:delay} shows the time it takes to detect traffic of interest. Solutions
that do not require iterative refinement have a delay that is equal to one window interval. 
{\em Fixed-Refinement} solutions require eight window intervals to detect traffic of interest. 
In contrast, \system learns better query plans---not only requiring lesser state in the data
plane but also detects events of interest sooner than {\em Fixed-Refinement} solutions.

To understand how sensitive our results are with respect to to parameters, like
threshold, sketch accuracy, and window intervals for each query, we ran the experiments 
varying each of these parameters. We observed that gains from iterative refinement are
higher for high threshold values and they diminish as the fraction that satisfies the query 
increases for lower threshold values. Similarly, for queries that require higher sketch 
accuracy, the performance gains are higher for \system and 
the performance of {\em Part-PISA} solutions are comparable to \system for lower sketch
accuracy---as they require less state to execute the queries at the finest refinement levels. 
The performance trends were unaffected by the choice of window intervals. However, longer 
window intervals require more state in the data plane.

\subsection{Benefits of Query Planning}
We now demonstrate that \system's ability to facilitate workload-driven query
planning makes efficient usage of available resources. 
Specifically, we show how \system's
query planning algorithm learns different plans under different system constraints.
We exhaustively explore the configuration
space by determining query plans for each configuration. For a given 
query and workload, we first determine the maximum amount of state ($B_{max}$) 
and the maximum number of tuples ($N_{max}$) required to execute all possible 
candidate query plans. We then evaluate a given query plan for all possible
configurations---varying system constraints from $[1 \dots 2 N_{max}]$ 
and $[1 \dots 2 B_{max}]$, respectively, in steps of size $20$ for each constraint. 
%\mcnote{Why is there a 2 times the parameter?}
%\ag{for exhaustive coverage of config space}

\begin{figure}[t!]
\centering
\includegraphics[width=\linewidth]{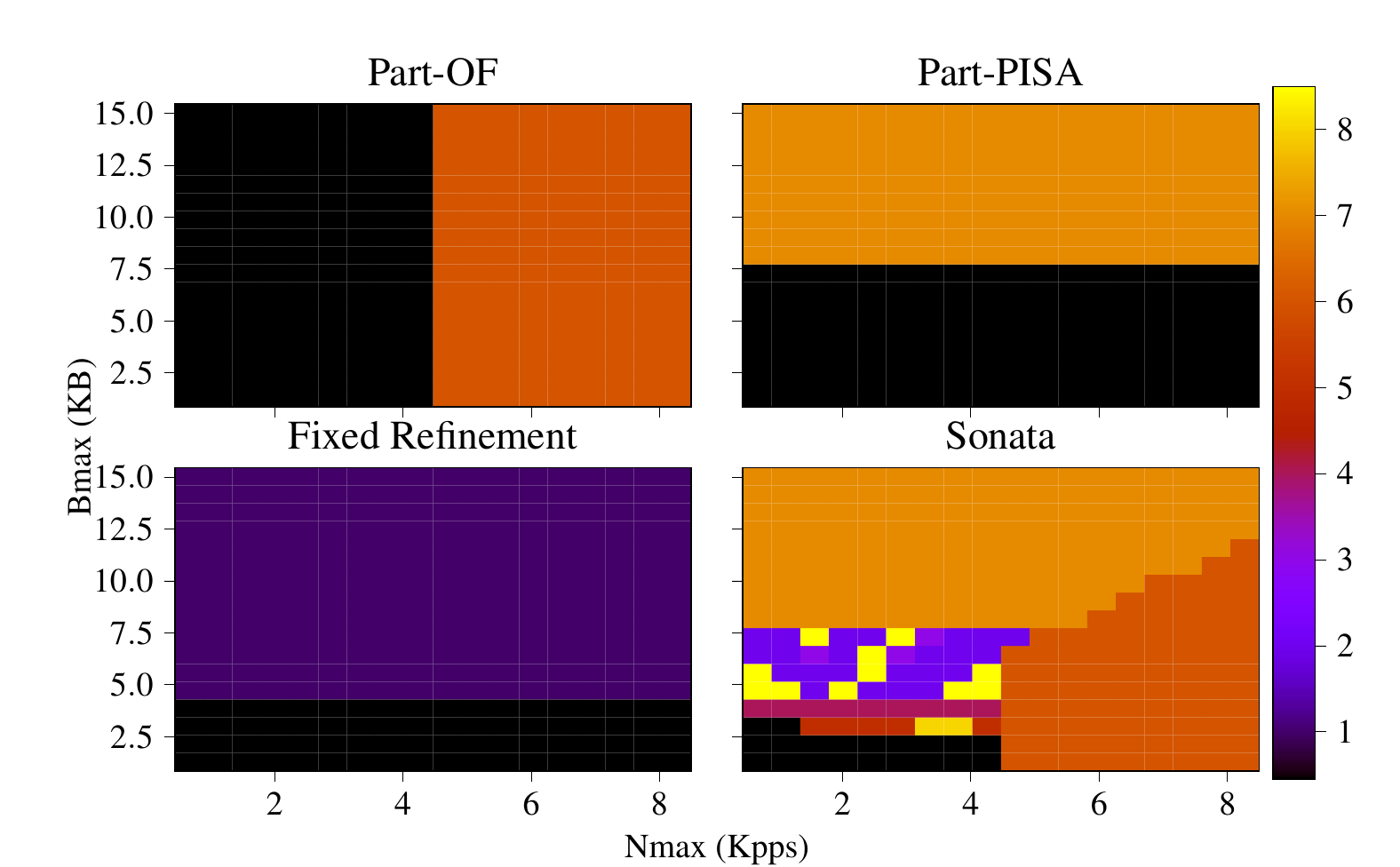}
\caption{Unique paths selected for various system configurations for {\em DDoS-UDP}
query. \system makes best usage of available resources selecting eight unique
query plans for different configurations.} 
\label{fig:alpha} 
\end{figure}

%\smartparagraph{Finding feasible query plans under constraints.}
Figure~\ref{fig:alpha} shows unique query plans selected for each combination of
system constraints for the {\em DDoS-UDP} query with the large IXP trace as workload. 
The black region in the graph shows configuration where no solution is possible, \ie~either 
the number of tuples sent to the stream processor exceeds the limit $N_{max}$ or the number of 
bits required in the data plane exceeds the limit $B_{max}$. Since {\em Part-OF} solutions
cannot execute stateful operations in the data plane, no solutions exists if $N_{max}$ 
is configured to be less than the number of incoming UDP packets. For 
{\em Part-PISA} solutions, no feasible query plans exist if the number of bits is configured
lower than what is required for executing both the {\tt distinct} and {\tt reduce} 
operations for {\em DDoS-UDP} in the data plane. {\em Fixed-Refinement} solutions
show a similar trend but require less state compared to {\em Part-PISA} solutions.
Finally, \system is able to learn the minimum cost query plans, finding feasible 
solutions for relatively more combinations of system configurations.
We observed similar results for the other queries and with the large ISP trace as 
workload. 

Table~\ref{tab:traininig} shows the total number of unique query plans observed for each query. 
We also observe that different queries (\eg, {\em DDoS-UDP} \& {\em SSpreader} or
{\em DDoS-UDP} \& {\em PortScan}) can result in different minimum cost query plans 
for the same system constraints which highlights the importance of selecting workload-driven 
query plans.

\begin{table}[ht!]
\begin{footnotesize}
\begin{center}
%\begin{tabular} {|l|p{0.75 in}|p{0.75 in}|p{0.75 in}|p{1 in}|p{1 in}|p{1 in}|}
\begin{tabular} {|c|cccc|}
%& \multicolumn{3}{c|}{\textbf{Partitioning}} & \multicolumn{2}{c|}{\textbf{Refinement}}  &\\ 
\hline
Network & DDoS-UDP & SSpreader & PortScan & All  \\ \hline
Large IXP & 8 & 5 & 5 & 17  \\
Large ISP & 6 & 4 & 13 & 20  \\
\hline
\end{tabular}
\end{center}
\end{footnotesize}
\caption{Total number of unique query plans selected by the runtime for the two real-world workloads.}
\label{tab:traininig} 
\end{table}

\subsection{Overheads}
We now quantify various overheads in selecting the query plans and updating
the queries for iterative refinement. 
\begin{figure}[t!] 
\centering
\includegraphics[scale=0.8]{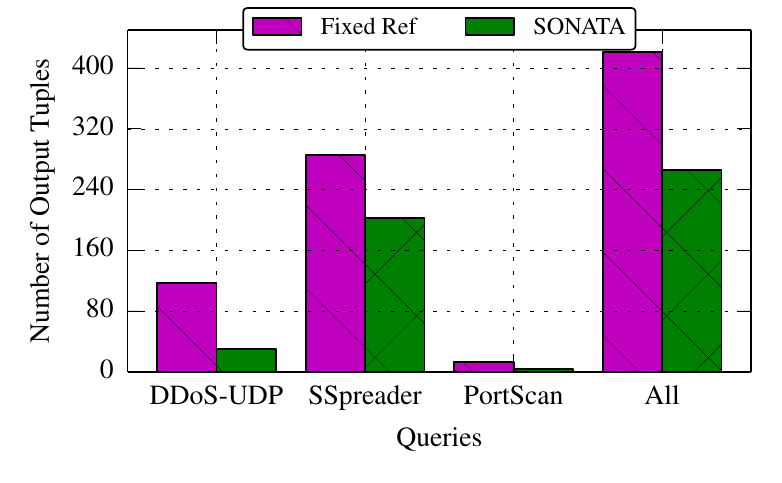}
\caption{Number of updates. \system requires fewer updates compared to 
{\em Fixed-Refinement} solutions.}
\label{fig:overheads} 
\end{figure}

\smartparagraph{Training Overheads.} 
To quantify the overhead for the runtime system to select the best query plan for a 
given query, we first determine the training duration ($M$), \ie,~the number
of window intervals required to accurately learn the best query query plan. 
To this end, we compute the training and test errors as we gradually increase the
number of window intervals used for training. These errors 
typically converge within 10--20 window intervals.
We observed that it takes 3-5 minutes to generate query plan graphs from 
the collected data of duration 20 window intervals, and 20--30 seconds to process 
the generated graphs to select query plans on our testbed.

\smartparagraph{Updating Query Executions.}
To quantify the time it takes to update the queries at the end of every window
interval for iterative refinement, Figure~\ref{fig:overheads} shows the number 
of output tuples at the end of every window interval for iterative refinement. 
Using our testbed setup, we quantify the time it takes for 
(1)~the runtime to process these output tuples, 
and (2)~the data plane driver to reset the hash tables for stateful operators, and
update the entries for the {\tt filter} tables. We observed that
the total overhead is around 100--150~ms. This high value is attributable to the
time
it takes to update the software switch which is around 100--120~ms. We expect this
number to 
be smaller for production-level switches like Tofino~\cite{tofino}---enabling
faster updates. 
%%\mcnote{Why do we expect this to be faster? The studies on OF switches always showed that real switch update performance is worse than that of software. Also you'd have to account for the network latency. So it does not seem realistic to expect lower delays here.}
%\ag{robh: not sure how to address this point, leaving it to you. }

\subsection{End-to-end Operation}

To demonstrate \system's end-2-end operation, we focus on the {\em Reflection Attack}
query and use the campus network trace as our workload. We load the packets 
from the trace and send them to the data plane target. Operating with a window
interval of one second, the query uses the refinement levels {\tt dIP/16} and {\tt dIP/32}. 
At time $t=20$  seconds, we inject additional packets at rate 300 packets/second, 
for a period of ten seconds. This injected (synthetic) traffic consists of multiple source 
addresses sending DNS packets of type {\tt RRSIG} to a single host. 

\begin{figure}[t!] 
\centering
\includegraphics[scale=0.8]{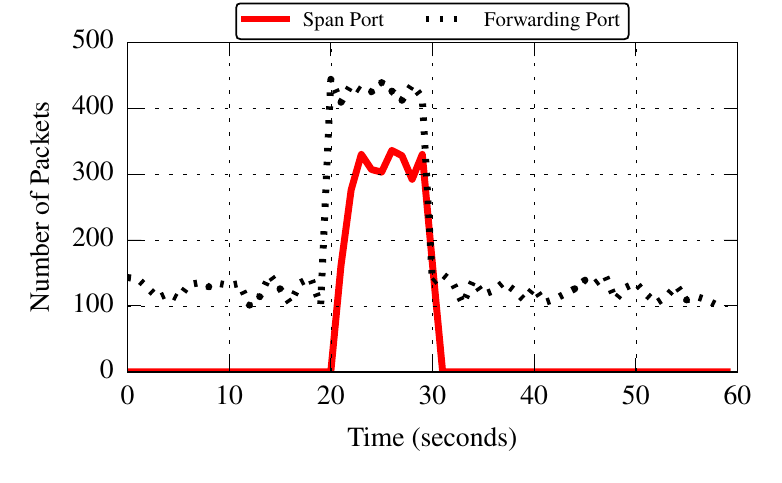}
\caption{End-to-end operation of \system for Query~\ref{payload-query}. Only the anomalous 
traffic is reported to the stream processor.}
\label{fig:macro} 
\end{figure}

We measure the number of packets at the forwarding and the span ports. 
Figure~\ref{fig:macro} shows how \system applies dataflow operations over 
the incoming packet stream, ignoring the normal traffic most of the time. As 
soon as anomalous traffic is injected, at the end of first window interval, \ie~one
second, it updates the packet processing pipeline to report only the traffic that
satisfies Query~\ref{asymm-query} to the stream processor which in turn is applying 
dataflow operations over the packet's DNS header fields. 

%Notice that for
%DDoS-UDP, the number of packets is high and the number of updates is
%low, thus parsing the output from the data plane targets takes most
%of the processing time. 

%\begin{figure}[t!] 
%\includegraphics[width=.66\linewidth]{figures/var_thresh.pdf}
%\caption{Performance gains for varying threshold values}
%\label{fig:overheads} 
%\end{figure}

\section{Related Work} 
\label{sec:related} 

\smartparagraph{Stream processing of network data.}
The most closely related technologies to \system are those that support analysis
of a combination of network traffic sources such as packet capture and IPFIX to
support
performance or security applications. Both Deepfield~\cite{deepfield}
and Kentik~\cite{kentik} support complex queries that require fusing information
from both IPFIX and packet traces, as discussed in Section~\ref{ssec:analysis};
yet, these systems are still relatively rigid in terms of the types of queries that
they can support, because they rely on existing collection technologies (\eg, IPFIX).
For example, existing Deepfield deployments perform analysis based on sampled
IPFIX records, since the tools do not support iterative query refinement and must
collect all data {\em a priori}. Neither Deepfield nor Kentik support queries over
high data-rate packet-level traffic streams.
Other systems can process streaming network
data~\cite{carney2002monitoring, sullivan1996tribeca, progme,gigascope,
abadi2003aurora, abadi2005design,amann2014count}. Some of these systems, such
as Tribeca~\cite{sullivan1996tribeca}  and Gigascope~\cite{gigascope},
tailored for network packet traces;
Tigon~\cite {tigon} and
OpenSOC~\cite{opensoc} provide similar abstractions and rely on streaming
analytics technologies such as Apache Hadoop, Spark, and Hive.
Yet, these systems 
%JEN: not clear this is true...
%have a less flexible query language
%than \system (\eg, they do not perform custom packet parsing to support a tuple
%abstraction), and they 
do not work directly on network switches, and do not introduce
scalability techniques such as sketches and real-time iterative
refinement that allow \system to scale.

\smartparagraph{Query planning.}
The database community has explored query optimization 
extensively~\cite{polychroniou2014track, mullin1990optimal, graefe1987exodus, catalyst}. 
An early example of query partitioning can be found in the Gigascope
system~\cite{gigascope} where the technique is used to minimize the
data transfer from the capture card to 
the stream processor. Query partitioning has also been extensively 
explored in the database literature in the context of query optimization~\cite{polychroniou2014track,mullin1990optimal,graefe1987exodus,catalyst,telegraphCQ}. 
The idea of query partitioning has also been used for distributing and efficiently executing
queries in sensor networks of low-power devices, using in-network operators for filtering
and aggregation~\cite{Madden.OSDI02,tinydb,Srivastava.PODS05}.
Geo-distributed analytics systems such as 
Clarinet~\cite{clarinet} and Geode~\cite{vulimiri2015global} partition queries across
geographically distributed compute and storage clusters.
Similarly, large-scale tracing of distributed systems, as in Fay~\cite{fay}, also adopts
distributed query execution across the cluster of machines, optimizing for factors such as
early trace data aggregation and reduced network communication.

\smartparagraph{Iterative refinement.}  Iterative zoom-in to reduce
the load on the data plane has been explored in earlier work such as
Autofocus~\cite{autofocus}, ProgME~\cite{progme}, and
HHH~\cite{jose2011online}.  Yet, these efforts either do not apply to
streaming data (\ie, they require multiple passes over the
data~\cite{progme}, or they use static refinement plans (\eg, HHH
performs zooms-in one bit at a time). More recently, Gupta et
al.~\cite{sonata-hotnets} explored the idea of iterative query
refinement in combination with query partitioning, but their system
only considers data-plane targets with fixed-function chipsets and
requires network operators to manually choose the refinement and
partitioning plan. In contrast, \system is designed for streaming
data, also supports drivers for programmable data-plane targets, and
selects workload-driven query plans.

\smartparagraph{Sketches.} Sketches have long been an active area of
research for many years in the theoretical computer science
community~\cite{muthu-book}. More recently, networking researchers
have explored applications of sketches to network traffic monitoring
in next-generation switches~\cite{univmon,opensketch,hashpipe}.  In
\system{}, our focus is on automatically generating an efficient
sequence of sketches for realizing a specific query, rather than
supporting a universal sketching platform~\cite{univmon,opensketch} or
a point solution for a specific query~\cite{hashpipe}.

%better reference for Tribeca is:
%M. Sullivan and A. Heybey, 
%Tribeca: A system for managing large databases of network traffic
%Proc. USENIX ATC, 1998.

\label{lastpage}
\balance \section{Conclusion}\label{sec:conclusion}
\system makes it easier for network operators to express queries for a range of network
telemetry tasks without worrying about {\em how} and {\em where} these tasks get executed. 
Using realistic queries and workloads, we demonstrate that \system
selects optimal query plans that require less state in the data plane and
reduce data rates for the streaming analytics systems in comparison with various state-of-the-art systems.

\end{sloppypar}

%\vspace{-0.1in}
%\section*{Acknowledgments}
% Comments for people we need to acknowledge in the final version.

\pagebreak

\small
\setlength{\parskip}{-1pt}
\setlength{\itemsep}{-1pt}
% \footnotesize % SPACE
\balance\bibliography{paper,vighata}
\Urlmuskip=0mu plus 1mu\relax
\bibliographystyle{abbrv}
%\bibliographystyle{abbrvnat_noaddr} % SPACE
%\theendnotes % ENDNOTES
%\input{appendix}

\end{document}